# Preprint notes

**Title of the article:**

Emotion AI in Workplace Environments: A Case Study

**Authors:**

Joni-Roy Piispanen, Rebekah Rousi

**Notes:**

# Emotion AI in Workplace Environments: A Case Study

Joni-Roy Piispanen[1][0009-0003-7533-4128] and Rebekah Rousi[1][0000-0001-5771-3528]

[1] University of Vaasa, Vaasa 65200, Finland
`{joni.piispanen,rebekah.rousi}@uwasa.fi`

**Abstract.** Emotion AI is an emerging field of artificial intelligence, intended to be utilized by organizations to manage and monitor employees' emotional states, supporting employee wellbeing and organizational goals. The current paper presents a case study that took place in a Finnish research institute, in which research participants ($N = 11$) were interviewed about their experiences of working in an Emotion AI environment. Our findings indicate that employees have a positive predisposition towards wellbeing monitoring in the workplace when benefits are perceived firsthand. Concerns however, manifest even in settings where there is existing familiarity with the technology, how it operates, and who is conducting the data collection, these are discussed in the findings. We additionally note that employee concerns can be mitigated via robust organizational policies, transparency and open communication.

## 1   Introduction

The emerging field of Emotion Artificial Intelligence (Emotion AI), or data-driven technology that collects, analyzes and interprets emotional data is currently being rese[1]arched and developed to support future employee wellbeing. Recently, Emotion AI comprising sensor-based technologies, passively and continuously measuring employee wellbeing in real-time, have been utilized to unobtrusively measure engagement, productivity, job satisfaction, mood, stress, organizational relationships and outcomes [1-3]. Yet, this type of technology designed to identify and respond to emotional states does not come without its dark sides [4]. Employees can be vulnerable to technological interventions in workplace contexts, when considering heightened surveillance, privacy, coercive control, marginalization and biases, power asymmetries and lack of explicit consent [5-10]. In fact, recent regulations by the European Union specifically restrict the use of emotion recognition in critical public sectors [11]. This illustrates the necessity to elucidate employee experiences, perspectives, attitudes and concerns regarding Emotion AI in workplace environments.

Currently, there are substantial gaps in empirical investigations of human-centered experiences in Emotion AI and its associated data collection in the workplace. Previous research consists mostly of theoretical and conceptual level studies, utilizing speculative methods. Empirical studies of privacy experience are still in their infancy,

---

particularly in terms of experimental studies that probe how privacy experience manifests in context. Thus, we chose to focus on employee perspectives and experiences, as examining the phenomenological aspects of Emotion AI is imperative to understanding how the technology impacts employee wellbeing. Our aim is to contribute to a growing body of interdisciplinary research surrounding Emotion AI's ethical and social implications in workplace environments [9, 10, 12]. We argue that employee concerns over Emotion AI and data collection in workplace environments can be reconciled through establishing pathways for communication and building trust.

## 2   Methods

We conducted a case study on the practicalities of utilizing Emotion AI in workplace environments, based on a previous research project at a Finnish research institute. The project collected data from environmental sensors (air quality, sound levels, motion detectors), work equipment (virtual sensors, pressure-sensitive chairs, video cameras, keyboard and mouse trackers), and wearable devices (smartphone trackers, sports watches, HRV monitors). This data was combined with employees' self-reported emotional states via surveys in a longitudinal study to track workplace wellbeing. The aggregated data was used to monitor employee wellbeing and functioning. The data was subsequently made available to employees for self-reflection.

We leveraged two qualitative research methods: secondary observation and semi-structured interviews. The former consisted of reviewing accounts, reports, and representations created by the researchers in the previous research project. The latter consisted of interviewing research participants about their experiences of working in an Emotion AI workplace environment, and interviewing researchers to validate their observations and experiences. Two research questions (RQs) guided our research:

- RQ1: How do employees experience and perceive an Emotion AI workplace environment designed to support their wellbeing?
- RQ2: What are employees' experiences of privacy and trust in Emotion AI workplace environments?

We recruited participants from the previous project who were willing to express their perspectives and experiences regarding working in an experimental Emotion AI workplace environment. Participants who expressed interest in participating in our study were contacted via email to fill in a Webropol survey-based informed consent form. In compliance with the General Data Privacy Regulation (GDPR), this initial email contained the privacy notice and research notification. Once participants had given their consent, a timeframe for interviews was established. In total, 11 participants consented to participate in our study. Nine participants (P1-P9) were employees who had participated in the previous research project, while two participants (P10-P11) were researchers conducting the research project in question. All participants were employees in research institutes and worked in Finland. Four of the participants were female and seven were male. Ages ranged between 25 and 64, with the largest age bracket being 45 to 54.

We adopted a semi-structured interview approach to collect participant experiences [13]. The interviews lasted between 30 and 120 minutes. The interviews were conducted in Finnish to facilitate disclosure by participants and to provide a more comfortable experience. The interviews were subsequently transcribed verbatim by the interviewer and translated from Finnish into English by hand for analysis. The translations were verified with translation software for accuracy and revised when necessary. We designed the interview protocol to begin with general topics and questions to establish rapport with participants, delving into more specific and sensitive topics further on in the interviews to avoid influencing the participant's answers. We felt it was necessary to mitigate potential researcher bias, recall bias, response bias and social desirability bias, given the sensitive and emotionally charged subject matter. The interview transcripts were analyzed utilizing reflexive thematic analysis [14]. Firstly, the interview transcripts were open-coded by paying close attention to the language and meaning used by the participants. Secondly, the codes were further refined to resolve any ambiguity or disagreements regarding granularity. Thirdly, following an iterative process the codes were developed into successively higher-level themes. Through this process patterns and commonalities in participant experiences could be identified.

## 3   Results

Our results indicate that: 1) employees have a generally positive attitude towards improving wellbeing in workplace environments even through embedded data-intensive technologies; 2) employees have an appreciation and a positive predisposition towards wellbeing monitoring at the workplace, given that they also benefit from this process; 3) even employees who have previous familiarity with the technology, its operation and data collectors are concerned about their privacy; 4) existing trust in employers, researchers, data infrastructures and operational procedures are not sufficient by themselves to alleviate privacy concerns; 5) the objectives of data collection, anonymity and the nature of organizations involved in data collection are the primary factors affecting employee experiences; and 6) transparency and open communication build trust, positively impacting experiences of Emotion AI in workplace environments. A summary of the results is depicted in Table 1 with thematic codes and evidence from interviews.

Table 1 Summary of results indicated by thematic codes and evidence from interviews.

| Thematic Codes | Interview Questions | Interview Quotes |
|---|---|---|
| Perceived Benefits of Data Collection | If we are only thinking about the risks and benefits, what perspective would you arrive at? | Well, in terms of benefits, (…) understanding what is stressful and how it can be reduced, that is valuable, and why not only from the employee's point of view, but also from the employer's point of view. (P2) |
| | | The enormous benefit of it was that you started to learn, you could see your activity in those applications (…) In it, perhaps, one became more aware of one's own ability to work and function. (P5) |
| | | However, since it was for a good cause, I wouldn't have minded no matter what kind of sensors there were. (P9) |
| Concerns over Privacy and Data Use | Did you have any particular concerns in terms of data privacy or ethics, and if you did, what kinds? | I'm of course interested in the fact that my data doesn't go to the wrong parties, those who could possibly misuse it. (P3) <br> Hmm, well there are risks. They are much easier to come up with. There is always a question about whether exactly what is said in the notices is being done. (P4) <br> (…), but I think it is a big risk if it is used for some kind of control. (P5) <br> I was wary of how the information would be utilized. What kind of data is collected (…) These were my initial concerns. (P7) |
| Trust in Data Handling | What is your overview, if we are talking about data protection and the ethical aspects? | Well as far as I can see, at least it had been well considered and it was brought up several times. Particularly, in what way the data is collected, for what purpose and what kind of people have access to what information and how well that data can or cannot be connected to individuals. (P1) <br> Of course, since this is a research project and I myself am an employee there, I knew at what level of ethics and other things this would be treated. So, there was nothing to worry about. (P6) <br> Additionally, knowing the researchers or at least at an organizational level knowing how these issues are approached, you can have a trusting attitude. (P8) |
| Transparency and Communication of Procedures | What was your general idea of how well it was communicated to you, for example, what the purposes are and what is done with the data afterwards? What kind of notion did you have? | I think communication and informing are really important in that regard. Being open about challenges and that some things don't work. (P5) <br> We did everything we came up with. To make it transparent and intelligible. (P10) <br> During the project, we tried to take into account and invested into data protection and privacy-related issues. (…) In the early stages, the extra headache that this caused was sizable, however, through the value of the data and the trust of the participants, it paid back in the end. (P11) <br> Users expressed that it motivates them when you get answers. Considering the level of understanding the end user has when answering |

> questions was important. (…) You can't just give a robotic answer, like, look at the documentation on page so and so. (P11)

Participants in our study expressed a general appreciation for wellbeing monitoring at the workplace and experienced it as beneficial on both an individual level as well as organizational level. Employees expressed an appreciation for continually receiving data about their wellbeing, mood and efficiency. Participants even expressed preparedness for more extensive data collection, given that they have some access to the data that is collected and, in turn, benefit from the process. If the end goal is promoting wellbeing at work and sufficient anonymity is preserved, employees feel more comfortable with data collection at the workplace. Anonymity, awareness of data collection and its purposes appear to be the primary factors affecting employee experiences, attitudes and perceptions.

Concerns by employees were varied, even if the perceptions related to the acceptability of monitoring technologies at the workplace were positive. Participants in our study considered themselves privacy conscious, yet in their view do not actively preserve their privacy or keep up to date with technological advancements and best practices. In relation to data privacy, the acceptance of data collection, and its processing in a workplace environment, participants were mainly worried about the purposes of data collection and whether the employer was monitoring their activity with the aim of scrutinizing their work. However, with sufficient anonymity, transparency and explainability of data collection processes these concerns were mitigated.

The shared sentiment among both researchers and research participants was that effective communication is needed. The results show indications that transparency and open communication build trust, resulting in a positive impact on employee experience. Especially on the researcher's side, the relationship between communication and trust was emphasized. However, organizational cultures and relationships to data collectors also affect trust in a significant manner.

## 4 Discussion

Our findings indicate that employees appear highly interested in receiving data about their wellbeing at work, while also showing an appreciation for organizations monitoring and managing how employees fare in workplace environments. In this respect, recent literature has emphasized employee opportunities to opt-out of workplace monitoring, adequate opportunities for exercising employee's rights, and the actual distribution of benefits from Emotion AI as contentious aspects [9-10, 12, 15-16]. Organizations planning to utilize Emotion AI need to consider transparency, communication and the purposes of data collection, when attempting to convince employees of the benefits. Additionally, geographical and cultural differences should be considered [17-18].

There were some limitations to this study that should be considered when evaluating the results. Participants in our study were employed by research institutes involved in the previous research project and indicated existing trust towards institutions and the research institutes conducting the research project. Additionally, participants in the previous research project volunteered to work in the experimental Emotion AI workplace environment, which could indicate a positive predisposition towards wellbeing monitoring and data collection. Furthermore, the research institutes in question are located in Finland and have an excellent reputation within the country both for their research quality as well as ethics and security standards. Moreover, all participants expressed knowledge of the European General Data Protection Regulation and at least a passing familiarity with organizational policies at the organization conducting the previous research project. As such, attitudes towards researchers, participation and data collection might be biased.

Further research is required to establish and validate commonalities in employee experiences of Emotion AI in workplace environments. Additionally, the relationship between age, gender and privacy-related behavior requires more comprehensive analysis.